\documentclass[two columns, amsmath,amssymb,aps, pra, reprint, groupedaddress]{revtex4-1}
\usepackage{graphicx}
\usepackage{dcolumn}
\usepackage{bm}

\begin{document}

\title{Estimation of gravitational acceleration with quantum optical interferometers}
\author{S. Y. Chen$^{1,2}$ and T. C. Ralph$^2$}
\email{ralph@physics.uq.edu.au} 
\affiliation{1.Quantum Institute for Light and Atoms, Department of Physics, East China Normal University, Shanghai 200062, China\\
2.Centre for Quantum Computation and Communication Technology, School of Mathematics and Physics, University of Queensland, Brisbane, Queensland 4072, Australia}

\date{\today}

\begin{abstract}
The precise estimation of the gravitational acceleration is important for various disciplines. We consider making such an estimation using quantum optics. A Mach-Zehnder interferometer in an ``optical fountain" type arrangement is considered and used to define a standard quantum limit for estimating the gravitational acceleration.
We use an approach based on quantum field theory on a curved, Schwarzschild metric background to calculate the coupling between the gravitational field and the optical signal. The analysis is extended to include the injection of a squeezed vacuum to the Mach-Zehnder arrangement and also to consider an active, two-mode SU(1,1) interferometer in a similar arrangement. When detection loss is larger than $8\%$, the SU(1,1) interferometer shows an advantage over the MZ interferometer with single-mode squeezing input. The proposed system is based on current technology and could be used to examine the intersection of quantum theory and general relativity as well as for possible applications.
\end{abstract}

\maketitle

\section{Introduction}

Precise measurement of Earth's gravity is of practical importance for many fields, such as navigation, geophysics and nature resource exploration. Such measurements can also be used as tests of general relativity  \cite{li}. Of considerable interest from both the fundamental and applied point of view is the coupling of quantum systems to gravity, for example as demonstrated by Colella, Overhauser and Werner \cite{cole}. Several proposals have been made to extend these types of experiments \cite{mag2, mag3,mag4, lui} and applications suggested such as enhanced global positioning systems and telecommunications \cite{sch2, man}.

Atomic system has been applied to the detection of Earth's gravity with high precision \cite{chu,sna, pet,pet1,mats,rosi,poli,rosi2,  gio, FS, chaohong}. In particular atomic fountains can achieve a precision of $\Delta g\approx 3\times10^{-9}g$ \cite{pet,chu,pet1}. Such systems may approach quantum limits for parameter estimation \cite{FS, chaohong, gio}. Some authors have considered the measurement of gravity via quantum optical methods \cite{kazuo, mag1, david, kish1, christ, kish2}. 
In order to determine quantum limits for the estimation of the gravitational field strength via such methods it is important to delineate the signal photons, which acquire a differential phase shift due to the gravitational field, from reference photons that are used to perform homodyne detection or to pump active media, etc.

In this paper, we analyse a quantum optical interferometer in an ``optical fountain" arrangement whereby a signal field is generated at some height, sent to a greater height where it is delayed, then returned to the original source height and interfered. The advantage of this arrangement is that all reference beams and optical pump fields remain at the source height and so do not acquire a signal. Hence the reference and pump fields can be consistently treated as ``free-resourses", whilst the signal photons are treated as the quantum resources. In this way we define a quantum standard limit (SQL) for quantum metrology of the gravitational field strength, as a function of the photon number in the signal beam, based on a standard Mach-Zehnder (MZ) type arrangement. We then examine surpassing this quantum limit using single mode squeezing in the MZ arrangement and by employing an active SU(1,1) type interferometer \cite{math, math2, boyd, pau,pau2}. The effects of internal and detection losses are included. The calculations are carried out in a general relativistic way using quantum field theory techniques on a curved background \cite{bir}.

The paper is organized in the following way. In Section \ref{2} we describe our various set-ups and derive analytical expressions for the relative sensitivities for estimating the gravitational acceleration. Subsection \ref{A} reviews phase estimation in a MZ interferometer. Subsection \ref{B} calculates the phase shift acquired by the signal beam propagating in the Schwarzschild metric as a function of the Schwarzschild radius and hence derives the standard quantum limit for estimation of the gravitational acceleration parameter. Enhancement of the estimation via the inclusion of a squeezed vacuum input in the presence of losses is analysed in subsection \ref{C}. Subsection \ref{D} considers the two-mode SU(1,1) interferometer. In section \ref{3} performance of the various set-ups is analysed numerically and best strategies under different conditions are proposed. In section \ref{4} we conclude and summarise our results.

\section{Setup and theory}
\label{2}
\begin{figure}[htp]
 \centering
  \includegraphics[width=3.5in]{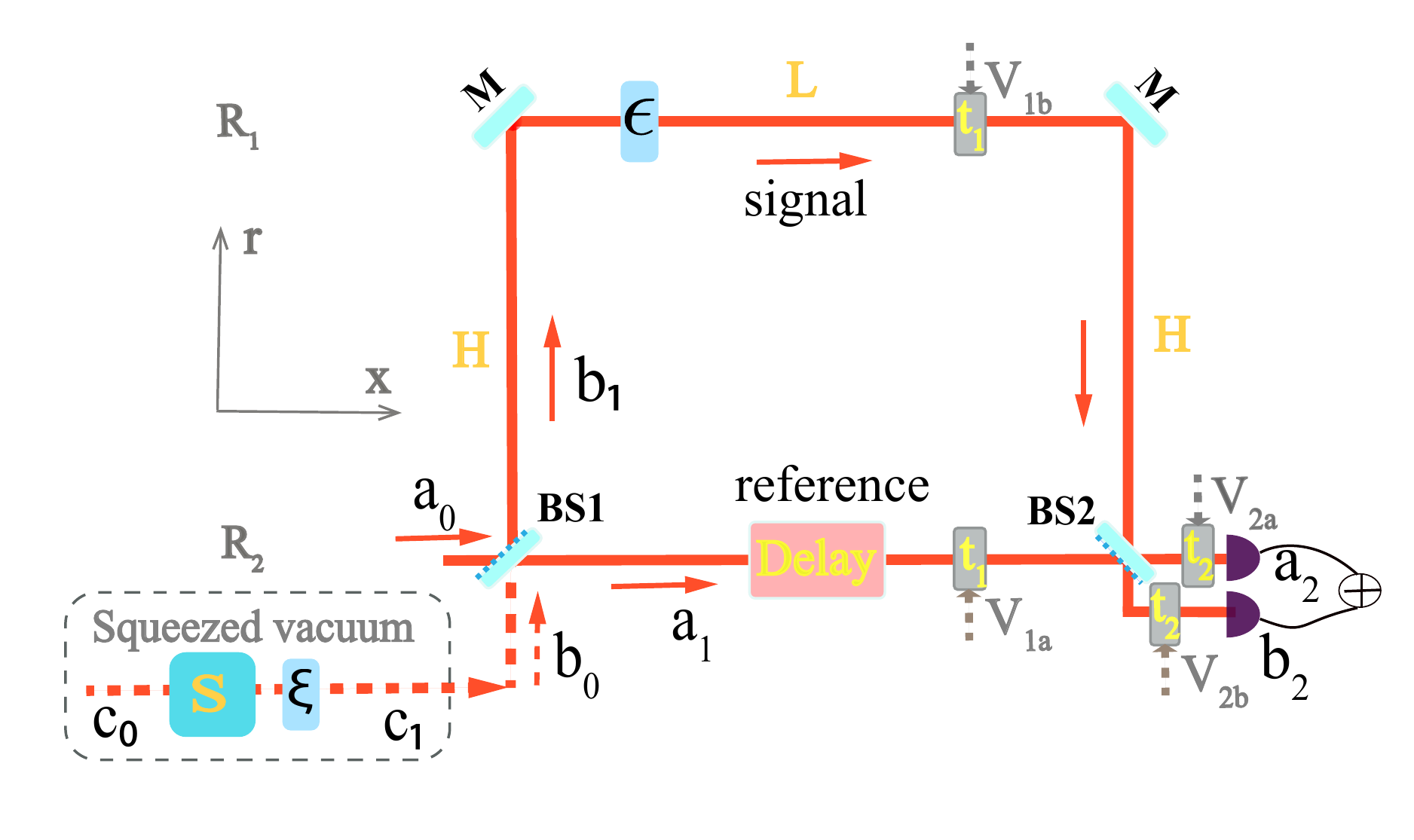}
  \caption{The optical interferometer is placed vertical to the ground from radial coordinate $R_2$ to coordinate $R_1$, with a local height difference $H$. For the losses, $t_1$ is the transmittance of the optical amplitude for internal loss, and $t_2$ is the transmittance for the external loss. The input states for BS1 are coherent light $a_0$ in one port and vacuum $b_0$ or squeezed vacuum $c_1$ in the other port. After BS1, the reference beam $a_1$ goes along the lower path and signal beam $b_1$ goes along the upper path. Finally, the outputs at the detection ports are $a_2$ and $b_2$. $\epsilon$ is the phase acquired by the signal beam of the interferometer. In the dashed rectangle is the single-mode squeezer. S: squeezer; $\xi$: phase modulator for single-mode squeezed vacuum; $c_0$: input vacuum field; M: mirror; $c_1$: single-mode squeezed vacuum.}
  \label{fig1}
\end{figure}

We propose the optical interferometer shown in Fig.\ref{fig1} to sense the local gravity. The interferometer is placed on the ground vertically along the radial direction of Earth. Due to space-time curvature, when the proper lengths of the two arms in the interferometer at different heights (upper path $b_1$ and lower path $a_1$) are the same, light travelling along the two arms will experience different local time, leading to different phases of the two optical paths. This phase difference can be detected by the intensity detection of the interferometer output.

In Fig.\ref{fig1}, a coherent field $a_0$ is injected into the first beam splitter BS1 while the other input port is injected with vacuum $b_0$, or the squeezed vacuum $c_1$ from the process in the dashed rectangle. After the BS1, the light travels along two paths. The signal beam $b_1$ travels vertically up to radius coordinate $R_1$, then reflected by mirror to travel along a horizontal path with a local distance $L$, finally reflected downwards vertically to the second beam splitter BS2 on the ground.  Meanwhile, the reference beam $a_1$ travels horizontally along the lower path on ground. After a time delay device, such as a laser delay line \cite{don, glo}, reference beam $a_1$ arrives at the second beam splitter BS2 and combines with signal beam $b_1$. The final outputs for detection after BS2 are light fields $a_2$ and $b_2$. 
The total proper length of the upper signal path is $2H+L$, where $H$ is the vertical proper distance from $R1$ to $R2$, $L$ is the proper length of the upper horizontal arm. For simplicity, we assume the delay device makes the proper length of the lower path equal to the proper length of the upper path $2H+L$. 
The gray cubes with labels $t_1$, $t_2$ in Fig.\ref{fig1} represent internal and external amplitude loss respectively. For simplicity, the losses on both signal beam $b_1$ and reference beam $a_2$ are the same. The dashed rectangle labeled with ``Squeezed vacuum'' is single-mode squeezed vacuum production. Vacuum $c_0$ is injected into squeezer S, and a phase sensitive squeezed single-mode vacuum field $c_1$ is produced, with a phase modulator $\xi$. Single-mode squeezed vacuum $c_1=b_0$ can be injected into BS1 in the vertical interferometer to decrease the detection noise and improve the sensitivity  \cite{kish1}. 

\subsection{Interferometer theory}
\label{A}
 The intensity transmittance of beam splitters BS1 and BS2 in the interferometer is $T$ and and reflectivity is $1-T$. With the coherent state $\hat a_0$ and vacuum state $\hat b_0$ as inputs, the input-output relations for the interferometer are,
\begin{eqnarray}
\hat a_2=&\left(T-(1-T)e^{i\epsilon}\right)\hat a_0+\sqrt{T(1-T)}(1+e^{i\epsilon})\hat b_0 \\
\hat b_2=&-\sqrt{T(1-T)}(1+e^{i\epsilon})\hat a_0 + \left(Te^{i\epsilon}-(1-T)\right)\hat b_0
\label{a2b2}
\end{eqnarray}
Here we set $\epsilon=\epsilon_0+\epsilon_G$ with $\epsilon_0=\pi$ a constant off-set, and if the gravity induced phase $\epsilon_G \ll \pi$, the outputs $a_2$ and $b_2$ in Eq.\ref{a2b2} can be approximated as 
\begin{eqnarray}
\hat a_2=&(1+(1-T)i\epsilon_G)a_0-\sqrt{T(1-T)}i\epsilon_G b_0\\
\hat b_2=&\sqrt{T(1-T)}i \epsilon_G  a_0-(Ti\epsilon_G+1)b_0 
\label{b2}
\end{eqnarray}

From homodyne detections at the outputs $a_2$ and $b_2$, the phase quadrature of both outputs $\hat X_a^{-}=i(\hat a_2-\hat a_2^{\dag}), \hat X_b^{-}=i(\hat b_2-\hat b_2^{\dag})$ are,

\begin{eqnarray}
\begin{split}
\hat X_a^{-}=&\sqrt{T(1-T)}\epsilon_G \hat X(b_0)-(1-T)\epsilon_G \hat X(a_0)\\
&+\hat Y(a_0) \\
\hat X_b^{-}=&-\sqrt{T(1-T)}\epsilon_G \hat X(a_0)+T\epsilon_G \hat X(b_0)-\hat Y(b_0) 
\end{split}
\end{eqnarray}

Here $\hat X(b_0)=\hat b_0+\hat b_0^{\dag}, \hat Y(b_0)=i(\hat b_0-\hat b_0^{\dag})$, $\hat X(a_0)=\hat a_0+\hat a_0^{\dag}, \hat Y(a_0)=i(\hat a_0-\hat a_0^{\dag})$. Focusing on the output $b_2$, the expectation and variance of quadrature $\hat X_b$ are
\begin{eqnarray}
\left\langle \hat X_b^{-} \right\rangle =& -2\sqrt{T(1-T)}\epsilon_G |\alpha|\\
\left\langle \Delta (\hat X_b^{-})^2 \right\rangle =& T(1-T)\epsilon_G^2+T^2\epsilon_G^2+1\approx 1
\label{dXb}
\end{eqnarray}

So we get the phase sensitivity of single output detection of $X_b^{-}$:
\begin{eqnarray}
\left(\Delta \epsilon_G \right)_b=\dfrac{\left\langle \Delta (\hat X_b^{-})^2 \right\rangle}{\dfrac{d\left\langle \hat X_b^{-} \right\rangle}{d\epsilon_G}}=\dfrac{1}{\sqrt{T}2\sqrt{N_{sig}}}
\label{dtheta}
\end{eqnarray}
Here $N_{sig}=(1-T)N_0$  is the photon number of the signal beam $b_1$ which senses the phase change through the upper path and $N_0=|\alpha|^2$ is the initial input photon number of the coherent state $a_0$.
The phase sensitivity of single output detection of the interferometer is the best $\left(\Delta \epsilon_G\right) _{b}=\dfrac{1}{2\sqrt{N_{sig}}}$ when $T \approx 1$. $T\approx 1$ means the photon number of the reference beam is much larger than the signal beam, which is an unbalanced interferometer, and also analogous to a homodyne detection apparatus with much stronger local oscillator field \cite{bri}. 

Another choice is to take joint homodyne detections of both outputs: $\hat X_j=\hat X_a^{-}+\hat X_b^{-}$ \cite{cave}. Under the same phase situation, we get the phase sensitivity of the interferometer from joint quadrature $\hat X_j$:
\begin{eqnarray}
\left(\Delta \epsilon_G\right)_j=\dfrac{\left\langle \Delta \hat  X_j^2 \right\rangle}{\dfrac{d\left\langle \hat X_j \right\rangle}{d\epsilon_G}}=\dfrac{\sqrt{2}}{(\sqrt{T}+\sqrt{1-T})2\sqrt{N_{sig}}}
\label{dtheta_los}
\end{eqnarray}

When $T=\dfrac{1}{2}$, the phase sensitivity from the joint homodyne detection is the best: $\left(\Delta \epsilon\right) _{j}=\dfrac{1}{2\sqrt{N_{sig}}}$. Thus the optimal phase sensitivities from the two different methods of detections achieve the same result with particularly chosen transmittance value $T$. For simplicity, we focus on the single output detection to define the standard quantum limit (SQL) for the gravity signal in the next step. 

\subsection{Gravity phase and SQL definition}
\label{B}
According to the Schwarzschild solution to general relativity equations, the space time around a massive body can be approximately described by the metric \cite{har}
 \begin{equation}
 \begin{split}
 ds^2=&-(1-\dfrac{r_s}{r})dt^2+(1-\dfrac{r_s}{r})^{-1}dr^2\\
&+r^2(d\theta^2+sin^2\theta d\phi^2)
  \label{metric}
   \end{split}
\end{equation}
Here $t$ is the time coordinate as seen by a far-away clock, $r$ is the radial coordinate defined as the circumference at that radius divided by $2\pi$, $\theta$ is the colatitude, $\phi$ is the longitude and $r_s$ is the Schwarzschild radius of the Earth. We work in units $c=1$,  where $c$ is the speed of light. In standard units, $r_s=\dfrac{2GM}{c^2}$, where $G$ is the gravitational constant. Here we neglect the rotation and ellipticity of Earth.

In the interferometer, the proper lengths of the upper signal path and lower reference path are set the same $2H+L$. The observer is assumed standing on the ground beside the interferometer. Because the horizontal size of the interferometer $L$ is much smaller than the radius of Earth $R_e \approx 6.7\times 10^{6} m$, the deflection of geodesic line of light can be neglected and the horizontal paths can be thought of as a geodesic line with constant radius $r$. For simplicity and to avoid the influence of the longitude $\phi$ change, the interferometer is devised to sit with the horizontal arm along the latitude line. When light travels along the horizontal path, the change of the longitude is $d\phi=0$. The vertical paths are taken to be approximately radial.

The proper length of the vertical path for signal beam $b_1$ can be calculated according to the Schwarzschild metric in Eq.\ref{metric} by substituting $dt=0, d\theta=0, d\phi=0$ to obtain the metric function $ds^2=\dfrac{1}{1-\dfrac{r_s}{r}}dr^2$. Hence, we get the proper distance of the vertical path is
\begin{equation}
\begin{split}
H&=\int_{R_2}^{R_1} ds=\int_{R_2}^{R_1} \dfrac{1}{\sqrt{1-\dfrac{r_s}{r}}}dr\\
&\approx R_1-R_2+\dfrac{r_s}{2}ln\dfrac{R_1}{R_2}
\end{split}
\label{H}
\end{equation}
The Taylor series expansion $\dfrac{1}{\sqrt{1-\dfrac{r_s}{r}}}\approx 1+\dfrac{r_s}{2r} $ is applied here. So the proper length of the whole upper path for signal $b_1$ is $L+2H\approx L+2(R_1-R_2+\dfrac{r_s}{2}ln\dfrac{R_1}{R_2})$.

The observed travelling time of the signal $b_1$ along the upward vertical path and downward vertical path is the same. The coordinate time of signal $b_1$ travelling along one vertical path is given by assuming a null geodesic
\begin{eqnarray}
0=ds^2=-(1-\dfrac{r_s}{r})dt^2+\dfrac{1}{1-\dfrac{r_s}{r}}dr^2
\end{eqnarray}
So we get the coordinate time of the signal travelling along the vertical arm from $R_2$ to $R_1$,
\begin{equation}
\begin{split}
t_{v}&=\int dt=\int_{R_2}^{R_1}\dfrac{1}{1-\dfrac{r_s}{r}}dr \approx R_1-R_2+r_s ln\dfrac{R_1}{R_2}
\end{split}
\label{t_v}
\end{equation}

For the signal $b_1$ travelling along the upper horizontal path $L$ at coordinate $R_1$, we derive
\begin{eqnarray}
0=ds^2=-(1-\dfrac{r_s}{r})dt^2+r^2d\theta^2
\label{t_h1}
\end{eqnarray}
Thus, we get the coordinate time of the signal beam $b_1$ horizontally travelling,
\begin{eqnarray}
t_{h}=\dfrac{L}{\sqrt{1-\dfrac{r_s}{R_1}}}
\label{t_h}
 \end{eqnarray}
Here we use $\int r d\theta\approx L$ with the approximation $sin \theta\approx \theta$, because the angle $\theta$ is small  with $L\ll R_e$. The local time observed by the observer on ground is 
\begin{equation}
\tau =\sqrt{(1-\dfrac{r_s}{R_2})} t
\label{local}
\end{equation}
 where $t$ is the coordinate time as seen by the far-away observer.
So, from the view of the observer on the ground, according to Eq.\ref{local}, the total local time of the signal beam $b_1$ travelling along the whole upper path is,
\begin{equation}
\begin{split}
\tau_{b}=\sqrt{(1-\dfrac{r_s}{R_2})}\left( \dfrac{L}{\sqrt{1-\dfrac{r_s}{R_1}}}+2(R_1-R_2+r_s ln\dfrac{R_1}{R_2})\right)
\end{split}
\label{tau_sig}
\end{equation}

For the reference beam $a_1$ on the ground, the travelling coordinate time along the lower path can be obtained from Eq.\ref{t_h1}. Because the proper length of the lower path is $2H+L$, the coordinate time for the reference beam is 
\begin{eqnarray}
t_{a}=\dfrac{L+2(R_1-R_2+\dfrac{r_s}{2}ln\dfrac{R_1}{R_2})}{\sqrt{1-\dfrac{r_s}{R_2}}}
\label{t_ref}
\end{eqnarray}
From Eq.\ref{local}, the local time for the reference beam $a_1$ by the observer on the ground  is
\begin{equation}
\begin{split}
\tau_{a}=L+2(R_1-R_2+\dfrac{r_s}{2}ln\dfrac{R_1}{R_2})
\end{split}
\label{tau_ref}
\end{equation}

The local time difference between the reference beam $a_1$ and the signal beam $b_1$ along different paths inside the interferometer is obtained from Eq.\ref{tau_sig} and Eq.\ref{tau_ref},
\begin{equation}
\begin{split}
\Delta \tau &= \tau_{a}-\tau_{b}=\dfrac{r_s}{2}(\dfrac{\Delta R L}{R_1R_2}+\dfrac{\Delta R^2}{R_2^2})\\
\end{split}
\label{dtau}
\end{equation}
Here $\Delta R=R_1-R_2$ is the coordinate separation. 

From Eq.\ref{H}, $\Delta R=R_1-R_2=H-\dfrac{r_s}{2} ln\dfrac{R_1}{R_2}$, and substituting this into Eq.\ref{dtau},

\begin{equation}
\begin{split}
\Delta \tau &\approx \dfrac{r_s}{2}(\dfrac{\Delta R L}{R_2^2}+\dfrac{\Delta R^2}{R_2^2})\\
&=\dfrac{r_s}{2}(\dfrac{(H-\dfrac{r_s}{2} ln\dfrac{R_1}{R_2}) L}{R_2^2}+\dfrac{(H-\dfrac{r_s}{2} ln\dfrac{R_1}{R_2})^2}{R_2^2})\\
&\approx \dfrac{r_s}{2}(\dfrac{H L}{R_2^2}+\dfrac{H^2}{R_2^2})
\end{split}
\label{dtau3}
\end{equation}

Finally at BS2, the frequency of the $b_1$ and $a_1$  is the same $\omega$ for the observer, but the travelling time by each beam is not the same, so a gravitational induced phase shift appears between the two beams.

Supposing the phase of $a_1$ is $\left( k\left( L+2H\right) -\omega \tau_{a} \right)=0$, the gravitationally induced phase difference between $a_1$ and  $b_1$ is,
 \begin{equation}
\begin{split}
 \psi &=\omega \Delta \tau =\omega \dfrac{r_s}{2}(\dfrac{H L}{R_2^2}+\dfrac{H^2}{R_2^2})=\omega \dfrac{g}{c^3}(HL+H^2)
\end{split}
\label{dpsi}
 \end{equation}
 where $g=\dfrac{r_s}{2R_2^2}$ is the local gravitational acceleration at coordinate $R_2$.

So in the MZ interferometer theory, replacing $\epsilon_G$ with this specific gravitationally induced phase shift $\psi$ in Eq\ref{dpsi}, we can define the standard quantum limit (SQL) for the relative sensitivity in the estimate of the gravitational acceleration $\dfrac{\Delta g}{g}$.
According to 
$$\dfrac{\Delta g}{g}=\dfrac{\sqrt{\left\langle \Delta X_b^{-2} \right\rangle}}{\dfrac{d\left\langle X_b^- \right\rangle}{d \psi}  \dfrac{d \psi}{d g} g}$$
we find
\begin{eqnarray}
\left( \dfrac{\Delta g}{g}\right)_{SQL}=\dfrac{1}{2\sqrt{N_{sig}}\dfrac{g\omega(H^2+LH)}{c^3}}
\label{sql}
\end{eqnarray}

In this gravity measurement situation, we regard the photon number $N_{sig}$ acquiring the relativistic phase shift as the photon number that counts in the SQL. The reference beam $a_1$ and local oscillator beams remain at coordinate $R_2$, thus do not acquire any differential phase due to the gravitational field. Hence we consider them ``free'' resources and only count the photons in the signal beam $b_1$.

\subsection{Squeezing and loss}
\label{C}
To suppress the noise of the signal and improve the sensitivity of the interferometer, single-mode squeezed vacuum \cite{luca} $\hat c_1=(G\hat c_0+g\hat c_0^{\dag})e^{i\xi}$, as produced from the squeezing source in the dashed rectangle in Fig.\ref{fig1}, can be applied to inject into the vacuum input port $b_0$ in the MZ interferometer. Here $G=cosh(r)$, $g=sinh(r)$ and $G^2-g^2=1$  and $r$ is the squeezing parameter of single-mode squeezer S. $c_0$ is the input vacuum for the squeezer S, as in Fig.\ref{fig1}. $\xi$ is the phase for adjusting the squeezed vacuum $c_1$. In realistic experiments, losses from the internal optical paths and detections are unavoidable. So the internal and external intensity losses of the interferometer are analysed with beam splitter models here. The amplitude transmittance for the loss model is $t_{j}$, and loss rate is $\eta_{j}$, with $t_j^2+\eta_j^2=1$. ($j=1$ is for internal loss; $j=2$ is for external loss.)
The input-output relations of the interferometer are
\begin{equation}
\begin{split}
\hat a_2=&t_1t_2((1+(1-T)i\epsilon_G)\hat a_0-\sqrt{T(1-T)}i\epsilon_G \hat b_0)\\
&+\eta_1 t_2(\sqrt{T} \hat V_{1a}+\sqrt{1-T}\hat V_{1b})+\eta_2 \hat V_{2a} \\
\hat b_2=&t_1t_2(\sqrt{T(1-T)}i \epsilon_G  \hat a_0-(Ti\epsilon_G+1)\hat b_0)\\
&+\eta_1 t_2(\sqrt{1-T}\hat V_{1a}+\sqrt{T}\hat V_{1b})+\eta_2 \hat V_{2b}
\end{split}
\end{equation}
Here $V_{1a}, V_{1b}$ correspond to the vacuum noise from internal loss ($V_{1a}$ for the reference beam $a_1$ and $V_{1b}$ for signal $b_1$), and $V_{2a},V_{2b}$ correspond to the vacuum noise from external loss ($V_{2a}$ for reference beam $a_2$ and $V_{2b}$ for signal $b_2$), as in Fig.\ref{fig1}.  Internal losses for signal beam $b_1$ and reference beam $a_1$ are set the same with the transmittance $t_1$ to keep the two beams noise balanced, which can be realized by adjusting the reference beam loss rate. It is the same for the external losses.

Substituting squeezed vacuum $c_1$ into the vacuum input port $b_0$, and introducing the gravity phase $\psi$ in Eq.\ref{dpsi}, the relative sensitivity of $g$ with single-mode squeezed input and losses is,
\begin{equation}
\left(\dfrac{\Delta g}{g}\right)_{sq}=\dfrac{\sqrt{t_1^2t_2^2e^{-2r}+\eta_1^2 t_2^2+\eta_2^2}}{t_1t_22\sqrt{T}\sqrt{N_{sig}}\dfrac{g\omega(H^2+LH)}{c^3}}
\label{lsq}
\end{equation}
The photon number of the signal beam is still taken to be $N_{sig}\approx (1-T) N_0$, where we assumed $(1-T) N_0\gg g^2$. When $T\approx1$, the interferometer has the best sensitivity for $g$. From Eq.\ref{lsq}, we can find the input squeezed state can improve the relative sensitivity $(\dfrac{\Delta g}{g})_{sq}$, while the losses are detrimental for the sensitivity. When there is no squeezing input $r=0$ and losses are ignored $t_1=t_2=1$, we recover Eq.\ref{sql}. 

\subsection{Two-mode SU(1,1) interferometer}
\label{D}
\begin{figure}[htp]
 \centering
  \includegraphics[width=3.5in]{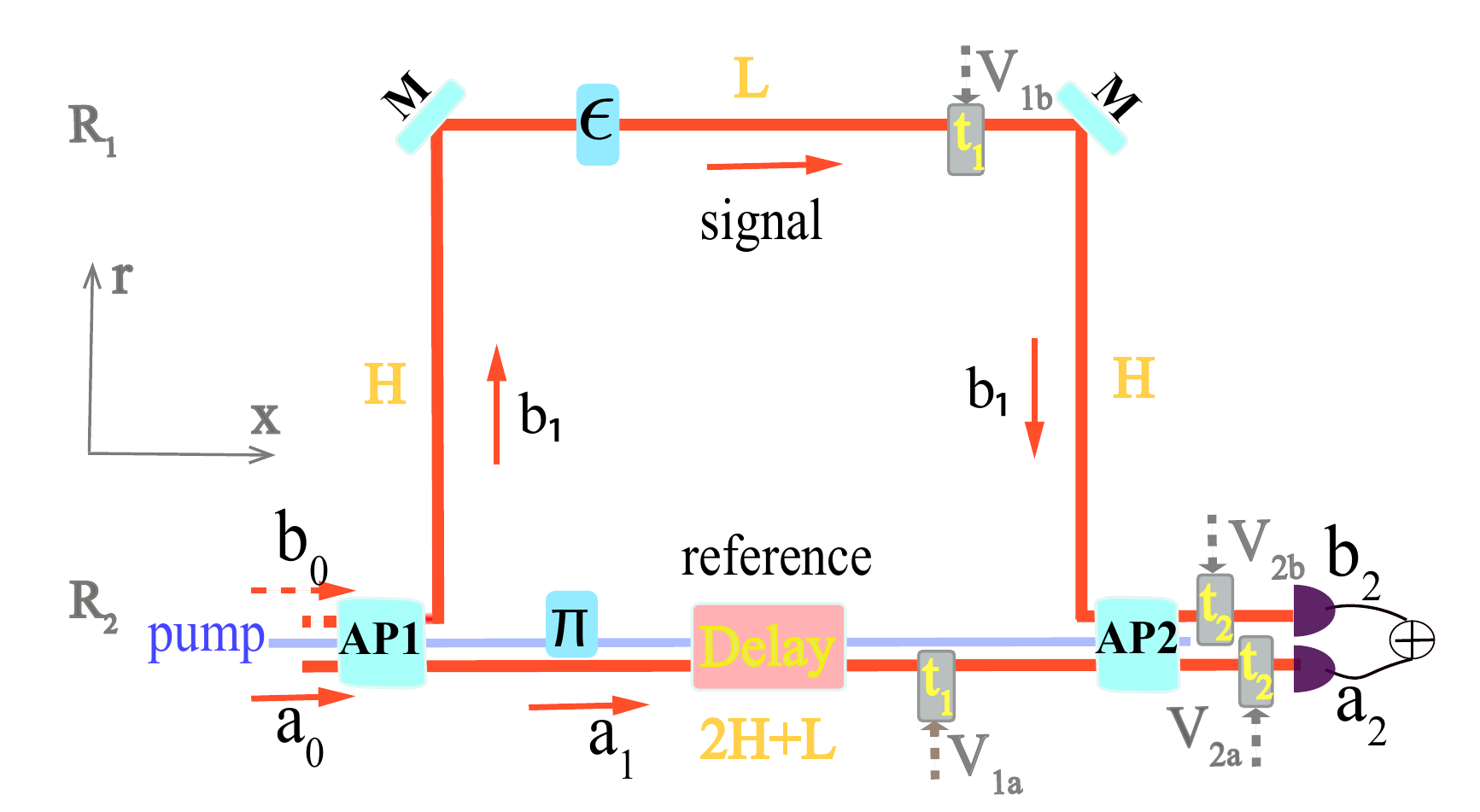}
  \caption{The optical interferometer is placed vertical to the ground from local height $R_2$ to local height $R_1$, with a proper length $H$. The input states are coherent light $a_0$ in one port and vacuum $b_0$ in the other port, aligned with pump. After the first amplification process (AP1), signal beam $b_1$ goes along the upper path and the reference beam $a_1$ goes  along the lower path.  And they combine at the second amplification process (AP2). Finally, the outputs at the detection ports are $a_2$ and $b_2$,  For the losses, $t_1$ is the internal  amplitude transmittance of the signal $b_1$ and reference $a_1$, and $t_2$ is the external amplitude transmittance for signal output $b_2$ and reference output $a_2$. Loss rates are $\eta_1=\sqrt{1-t_1^2}$ and $\eta_2=\sqrt{1-t_2^2}$. $V_{1a},V_{1b}, V_{2a}, V_{2b}$ are induced vacuum noise during the lossy channel.  A delay device is used to make the proper lengths of the upper path and lower path the same $2H+L$. } 
  \label{fig2}
\end{figure}

As suggested by some work \cite{pau,pau2,math,math2,boyd}, the two-mode SU(1,1) interferometer \cite{yurk, ou, bri} made up of two parametric amplification processes instead of two beam splitters, may remedy the sensitivity reduction from intensity losses with the advantage of two-mode squeezing. So the application of the SU(1,1) interferometer  for gravity field measurement is also analysed here.  We introduce a two-mode SU(1,1) interferometer as in Fig.\ref{fig2}, which is the analogue to the MZ interferometer in Fig.\ref{fig1}, where the signal beam $b_1$ goes along the upper path and the idler beam $a_1$ and the pump beam goes along the lower path as reference. The AP1 and AP2 in Fig.\ref{fig2} represent the two parametric amplification processes, which work with the input-output relation: $\hat a_{out}=G\hat a_{in}+g\hat b_{in}^{\dag}$, $\hat b_{out}=G\hat b_{in}+g\hat a_{in}^{\dag}$. $G$ is the gain of the amplitude, and $G^2-g^2=1$. A linear phase of $\pi$ is placed on the pump beam. A small gravitational phase shift, $\epsilon_G \ll \pi$ on the signal beam $b_1$ is again assumed.

The input-output relations of this scenario are 
\begin{equation}
\begin{split}
\hat a_2=&(G^- +g_1g_2i\epsilon_G)\hat a_0+(g^- +g_2G_1i\epsilon_G)\hat b_0^{\dag}\\
\hat b_2=&-(G^- +G_1G_2i\epsilon_G)\hat b_0-(g^- +G_2g_1i\epsilon_G)\hat a_0^{\dag}
\end{split}
\end{equation}
Here $G^-=G_1G_2-g_1g_2$, $g^-=G_2g_1-g_2G_1$.
The best relative sensitivity of gravity acceleration in the two-mode SU(1,1) interferometer in Fig.\ref{fig2} by single homodyne detection at output $b_2$ is
\begin{equation}
 \left(\dfrac{\Delta g}{g}\right)_{2sq}=\dfrac{\sqrt{t_1^2t_2^2 \dfrac{(g^-)^2+(G^-)^2}{G_2^2}+\eta_1^2 t_2^2(1+\dfrac{g_2^2}{G_2^2})+ \dfrac{\eta_2^2}{G_2^2}}}{2t_1 t_2\sqrt{N_{sig}-g_1^2}\dfrac{g\omega(H^2+LH)}{c^3}}
 \label{s1}
\end{equation}
The first and second amplification process have gains $G_{i}=cosh(r_{i})$ and $g_{i}=sinh(r_{i})$ with $G_i^2-g_i^2=1$ ($i=1, 2$).  $r_1,r_2$ are the squeezing parameters in the first and second amplifications in the SU(1,1) interferometer.  Here the photon number of the gravitational induced phase sensing signal beam is $N_{sig}=g_1^2(N_0+1)\approx g_1^2N_0$, where we have assumed the photon number from the coherent source is large, $N_0 \gg 1$. 

If the two gains of the two amplifications are set the same, $r_1=r_2$, we have a sensitivity with a simpler form
\begin{equation}
\left(\dfrac{\Delta g}{g}\right)_{2sq}=\dfrac{\sqrt{\dfrac{t_1^2t_2^2}{G_2^2}+\eta_1^2 t_2^2(1+\dfrac{g_2^2}{G_2^2})+\dfrac{\eta_2^2}{G_2^{2}} }}{2t_1 t_2\sqrt{N_{sig}}\dfrac{g\omega(H^2+LH)}{c^3}}
\label{s11}
\end{equation}

When the gains for both amplification processes are unity, $G_1=G_2=1$, the sensitivity is the same as the MZ interferometer in Eq.\ref{lsq} with no squeezing. If losses are also neglected, $t_1=t_2=1$, the gravity acceleration sensitivity reaches the SQL in Eq.\ref{sql}.

To avoid losing information, joint homodyne detection of $X_j=X_a+X_b$ \cite{pau, pau2, cave} using both outputs of the SU(1,1) interferometer is analysed. After calculation, the best relative sensitivity of gravity acceleration from the joint homodyne detection is 
\begin{equation}
 \left(\dfrac{\Delta g}{g}\right)_{2jq}=\dfrac{\sqrt{2 t_1^2t_2^2 e^{-2r_{1}}+ 2 \eta_1^2 t_2^2+ 2e^{-2r_{2}}\eta_2^2}}{2t_1 t_2\sqrt{N_{sig}} \dfrac{g\omega(H^2+LH)}{c^3}}
\label{su2}
\end{equation}
 
We compare this joint detection result in Eq.\ref{su2} to the single detection result in SU(1,1) interferometer in Eq.\ref{s1}. The joint detection result shows the squeezing parameter $2e^{-2r_1}$ for the first term in the numerator in Eq.\ref{su2}, while the single detection result Eq.\ref{s1} shows a factor $cosh^{-2}(r_2)$. These two functions cross at certain squeezing parameters, which will be shown in the next section.

We also compare this two-mode squeezing result Eq.\ref{su2} to the sensitivity with single-mode squeezing in the MZ interferometer as Eq.\ref{lsq}. The first and second terms on the numerator in Eq.\ref{su2} are twice as large as the MZ interferometer, with the common squeezing benefit term $e^{-2r_1}$ on the first term. For the third term in the numerator in Eq.\ref{su2}, the loss term $\eta_2$ in two-mode SU(1,1) interferometer joint detection is decreased by a factor of $2e^{-2r_2}$ than the MZ interferometer in Eq.\ref{lsq}. 

The best strategy between these three situations will be analysed using the numerical comparisons in the next section.

\section{Parameters Analysis}
\label{3}
The relations between $ \dfrac{\Delta g}{g}$ and different parameters in the optical interferometer are analysed here considering particular experimental conditions.

 \begin{figure}[htp]
 \centering
  \includegraphics[width=3.5in]{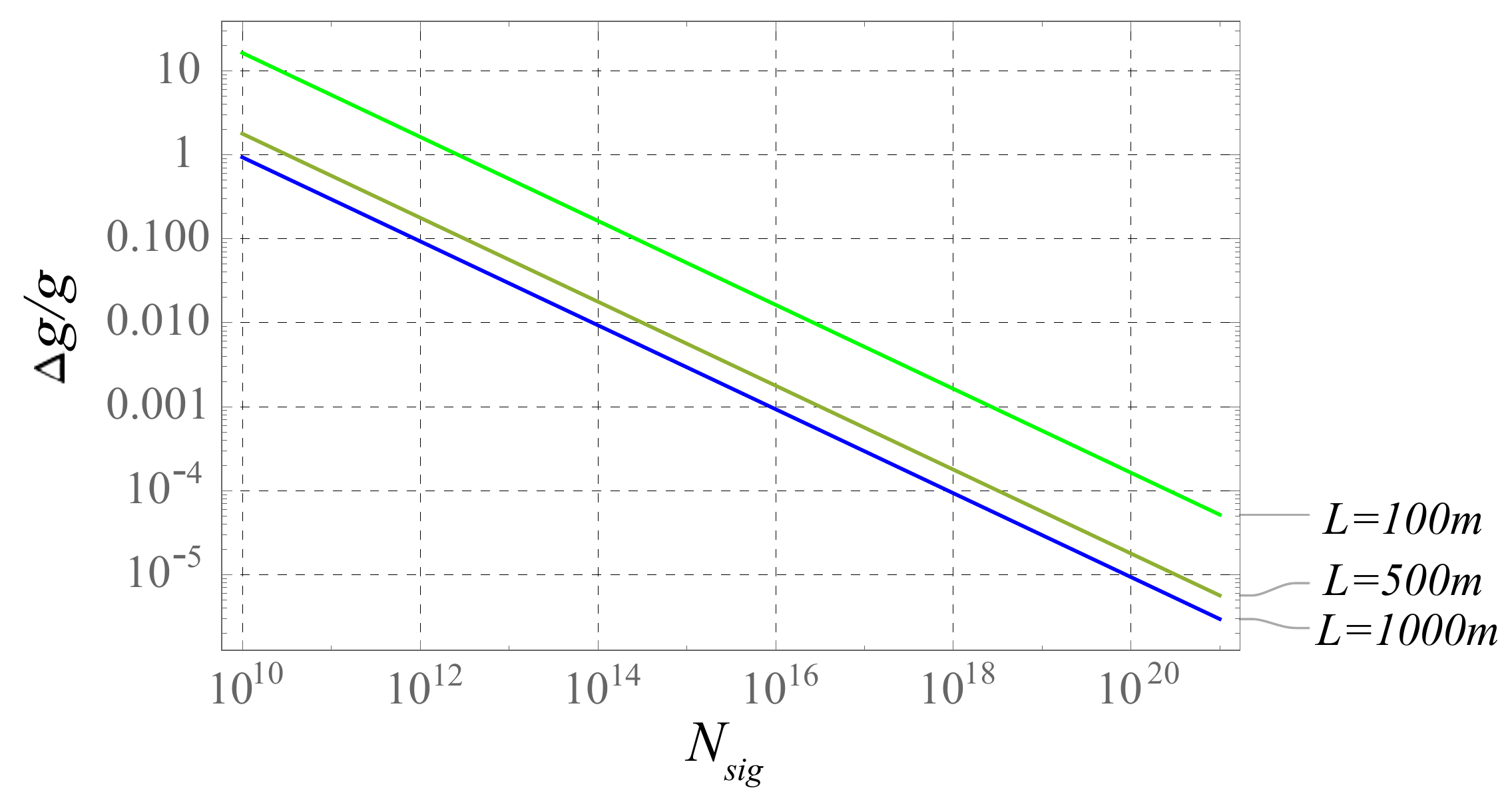}
  \caption{$\dfrac{\Delta g}{g}$ against $N_{sig}$. The proper length of upper horizontal arm is set $L=100m$ (green), $L=500m$ (cyan), $L=1000m$ (blue). 
The height of the vertical arms of the interferometer in Fig.\ref{fig1} is set $H=50m$. Input coherent light has the frequency $\omega=2.82\times 10^{14}Hz$ and works in continuous light mode. The reference gravity acceleration is $g=9.8m/s^2$, and the speed of light is $c=3\times 10^8m/s$. }
  \label{1gxn}
\end{figure}

First we estimate the required signal photon number $N_{sig}$ and upper horizontal arm length $L$ in the MZ interferometer according to Eq.\ref{sql} , to determine which parameters are good to choose when operating at the standard quantum limit (SQL). The result is shown as Fig.\ref{1gxn}. 		
The $x$-axis is the photon number of the signal beam $N_{sig}$ which experiences the gravity phase shift, and the $y$-axis is $\dfrac{\Delta g}{g}$. The three lines with different colors are corresponding to different lengths $L$ of the upper horizontal arm, with fixed vertical arm height of $H=50m$. Input coherent light is chosen with the wavelength of $\lambda=1064nm$, corresponding frequency $\omega=2.82\times10^{14}Hz$. We assume the interferometer is operated in continuous light mode. 
 
From the Fig.\ref{1gxn}, with a proper length of upper horizontal arm $L=100m$, the relative sensitivity of gravity acceleration can reach $\dfrac{\Delta g}{g}=5\times 10^{-3}$ with signal photon number $N_{sig}=10^{18}$, corresponding to a $1s$  detection with a continuous wave of power $1W$. When the  length  is $L=1000m$, the relative sensitivity can reach $\dfrac{\Delta g}{g}=10^{-4}$ under the same power conditions.  
 
  \begin{figure}[htp]
 \centering
  \includegraphics[width=3.5in]{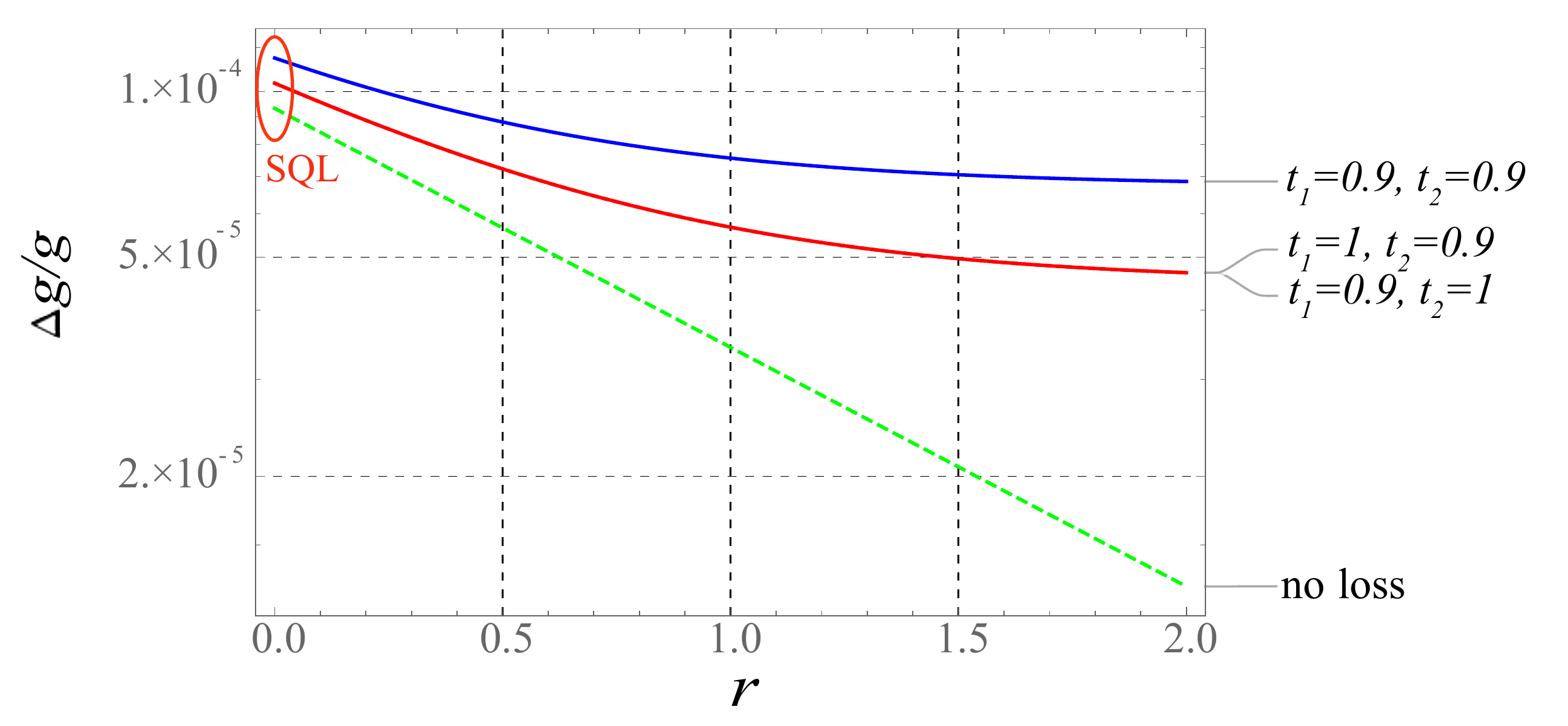}
  \caption{$\dfrac{\Delta g}{g}$ against the single-mode squeezing parameter $r$ of the squeezed input vacuum in the MZ interferometer.  Different lines are  sensitivities with different transmittance values $t_1$, $t_2$.   Other parameters are $H=50m, L=1000m$, $\omega=2.82\times 10^{14}Hz$, $c=3\times 10^8m/s$, $g=9.8m/s^2$ and signal beam photon number $N_{sig}=10^{18}$. }
  \label{2gtr}
\end{figure}
 
We now consider the inclusion of squeezing to improve the sensitivity and the detrimental effect of losses in the MZ interferometer. The relation between $\dfrac{\Delta g}{g}$ and the single-mode squeezing parameter $r$  is analysed in Fig.\ref{2gtr}. Different lines correspond to the different levels of internal and external losses.
The green dashed line is the sensitivity $\dfrac{\Delta g}{g}$ without losses, $t_1=t_2=1$. The red solid line is with internal transmittance $t_1=0.9$ and no external loss $t_2=1$, or only external loss $t_2=0.9$ and no internal loss $t_1=1$. According to Eq.\ref{lsq}, the effects on $\dfrac{\Delta g}{g}$ from internal loss and external loss are the same when only internal loss or only external loss exists. The blue line is corresponding to when both internal loss and external loss exist, $t_1=t_2=0.9$.  Squeezing improves sensitivity but its effectiveness is reduced by the losses.
The effective SQL with losses are given by the starting points at $r=0$ in different lines. 

\begin{figure}[htp]
 \centering
  \includegraphics[width=3.5in]{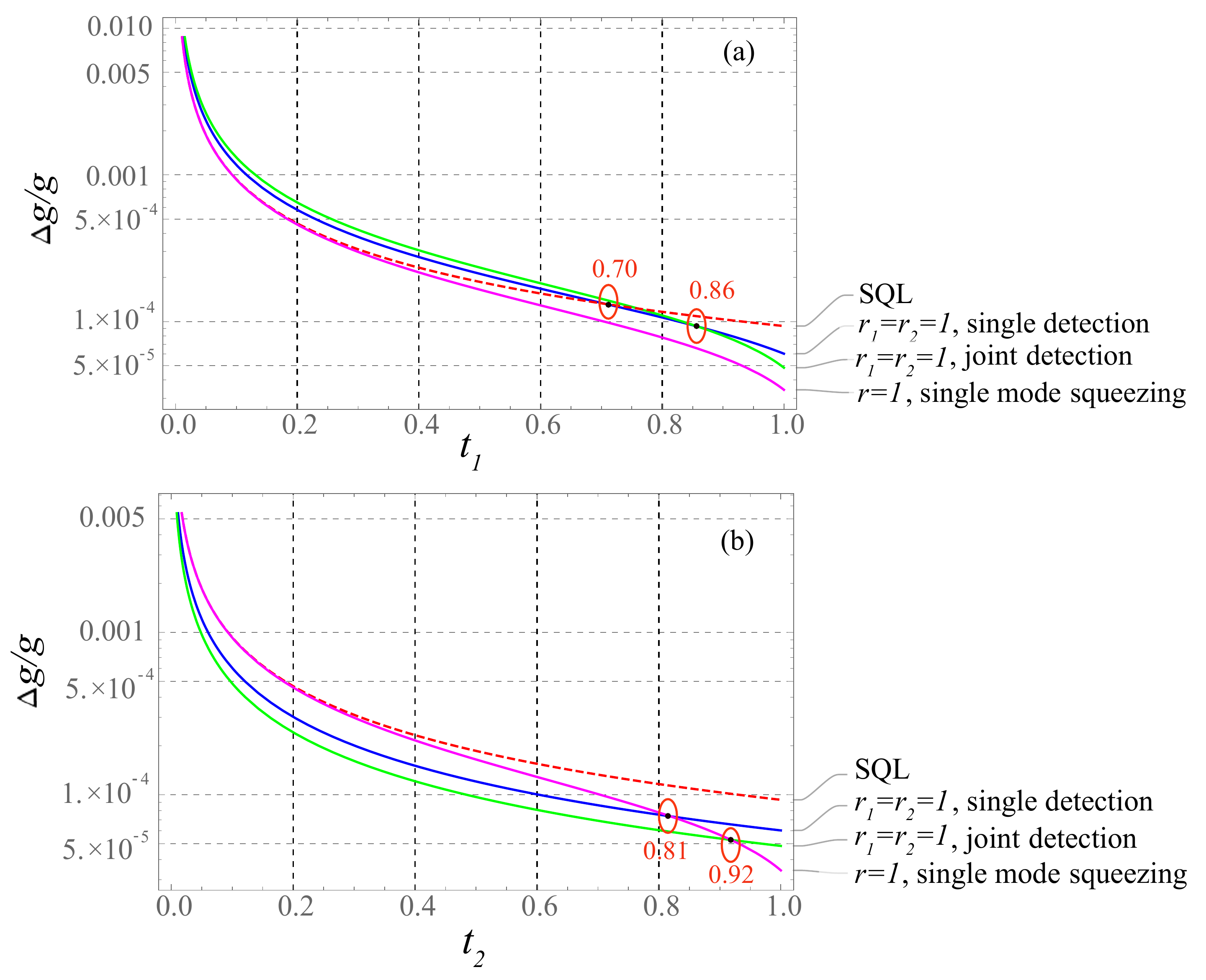}
  \caption{(a) $\dfrac{\Delta g}{g}$ against the internal transmittance $t_1$; (b) $\dfrac{\Delta g}{g}$ against the external transmittance $t_2$.  In either (a) or (b), the red dashed line represents the effective SQL. The blue and green lines are corresponding to the sensitivity of SU(1,1) interferometer with single homodyne detection (blue) and with joint homodyne detection (green) with $t_1=1$, and with squeezing parameters $r_1=r_2=1$; magenta solid line is corresponding to the sensitivity of MZ interferometer with single-mode squeezing parameter $r=1$. Other parameters are $H=50m, L=1000m$, $\omega=2.82\times 10^{14}Hz$, $c=3\times 10^8m/s$, $g=9.8m/s^2$ and signal beam photon number $N_{sig}=10^{18}$.  }
  \label{4grt2}
\end{figure}

Finally, the performance of the two-mode SU(1,1) interferometer with transmittance parameters $t_1,t_2$ and squeezing parameters $r_1, r_2$ are analysed in Fig.\ref{4grt2}.  

In Fig.\ref{4grt2}(a), the $x$-axis is the internal transmittance $t_1$, and the external transmittance is set perfect $t_2=1$. For the whole range of $t_1$, relative sensitivities $\dfrac{\Delta g}{g}$ from two-mode SU(1,1) interferometer (blue and green) are not as good as results from the MZ interferometer with single-mode squeezing input (magenta). When $t_1<0.70$, the $\dfrac{\Delta g}{g}$ from taking single detection or joint detection in the SU(1,1) interferometer are both not good enough to break the SQL.  For the comparison between the two different detection methods in SU(1,1) interferometer, when $t_1>0.85$, joint detection result (green) is lower than single detection result (blue). When $t_1<0.85$, single detection is better than joint detection.     

In Fig.\ref{4grt2}(b), the effect from external loss is analysed and shown. The $x$-axis is the external transmittance $t_2$, and the internal transmittance is set $t_1=1$. As for the whole range of $0\leq t_2\leq 1$, the results from two-mode SU(1,1) interferometer (blue and green) and from MZ interferometer with single-mode squeezing (magenta) are all better than the effective SQL (dashed red), which shows that the internal loss is more detrimental than the external loss for the sensitivity, compared to the Fig.\ref{4grt2}(a). When only external loss exists, the joint detection of SU(1,1) interferometer (blue) shows better sensitivity than the single detection (green).  When taking joint detection with $t_2<0.92$ or taking single detection with  lower than $t_2<0.81$ in the SU(1,1) interferometer, the sensitivity $\dfrac{\Delta g}{g}$ in SU(1,1) interferometer is better than the result in MZ interferometer with single-mode squeezing (magenta). But when external loss $t_2>0.92$, MZ interferometer with single-mode squeezing is the best.

In summary, when only internal loss exists, the SU(1,1) interferometer shows no advantage over the  MZ interferometer with single-mode squeezing input. Only when the external transmittance is $t_2<0.92$, the SU(1,1) interferometer has an advantage over the MZ interferometer with single-mode squeezing input. In the realistic measurement, both internal and external losses exist, so the best strategy depends on the conditions. It is currently quite possible to control the external loss within $10\%$, and the detector efficiency has been reported as good as $98\%$ \cite{rom,tit}. In that situation, the MZ interferometer with single-mode squeezing input and single output homodyne detection is easier to operate and has better sensitivity. 

\section{Conclusion}
\label{4}
We have discussed a system for precision measurement of the gravity acceleration constant $g$, based on the interferometry of optical fields interacting with the space time curvature of Earth.  The output of the interferometer is sensitive to the phase shift of the signal beam, which contains the information about the gravity. Thus such an optical interferometer can be used for gravity estimation. 

The photon number of the signal beam that can be carried up, and experience the gravity phase shift is  limited by current technology. So we define the standard quantum limit (SQL) of gravity acceleration $g$ based on the photon number limit of the signal beam without any loss or squeezing, $\left(\dfrac{\Delta g}{g}\right)_{SQL}=\dfrac{1}{2\sqrt{N_{sig}} \dfrac{g\omega(H^2+HL)}{c^3}}$. The resources operated on ground, such as the pump and reference beams, are regarded as ``free resources''.  
With single-mode squeezing input into the MZ interferometer, the relative sensitivity of $g$ on Earth can be measured with a better sensitivity than SQL. With a squeezing parameter $r=1$, the sensitivity $\dfrac{\Delta g}{g}$ can be improved by a factor of $\dfrac{1}{e}$ of magnitude. The effects from losses are also analysed.

Two-mode SU(1,1) interferometer is introduced to find the best strategy for the experiment when losses are inevitable. We show that with detection loss rate $\eta_2>8\%$, the two-mode SU(1,1) interferometer will show improvement for the sensitivity compared to  MZ interferometer with single-mode squeezing input. But for small external loss or the internal loss dominating, two-mode SU(1,1) interferometer has no advantage, and MZ interferometer with single-mode squeezing input is suggested as the first choice.

In principle, this system can be used for precision measurement of Earth's gravity. One could imagine deploying such a system between satellites and extending it to the gravity of other planets or celestial bodies. Sensitivity is improved by larger power of laser, shorter wavelength, larger size or longer probing time.

Comparing to the atomic system, in order to achieve the same level of sensitivity $\Delta g= 3 \times 10^{-9}g$, the optical system would need for example $N_{sig}=7.1 \times 10^{24}$ (Mega Watt power), $L=5km$, $r=1$, $H=50m$ for $1s$ detection. On the other hand, observation of the gravitationally induced phase shift in quantum optics does seem within the reach of current technology with this type of systems.

\begin{acknowledgments}
We acknowledge the support from the Australian Research Council Centre of Excellence for Quantum Computation and Communication Technology (Project No. CE170100012) and the China Scholarship Council programs. We also would like to extend our gratitude to the useful discussions with Magdalena Zych, Austin Lund, S.P. Kish and Marco Ho from the University of Queensland; and to Daiqin Su from Xanadu located in Toronto for plenty of mindful communications about this work.
\end{acknowledgments}

\bibliography{basename of .bib file}

\end{document}